\documentclass{optica-article}
\journal{opticajournal} 

\articletype{Research Article}

\usepackage{mathtools}
\usepackage{bm}
\usepackage{physics}
\usepackage{comment}

\usepackage{graphicx}
\usepackage{subcaption}

\usepackage{cleveref}
\crefname{section}{Section}{Sections}
\crefname{appendix}{Appendix}{Appendices}
\crefname{figure}{Figure}{Figures}
\crefname{table}{Table}{Tables}
\crefname{equation}{Eq.}{Eqs.}
\crefname{assumption}{Assumption}{Assumptions}

\usepackage{siunitx}

\newcommand{\mrm}[1]{\mathrm{#1}}

\begin{document}
\title{Dynamically-scaled local patch method for numerically simulating the scattering effects of small particles into large beams}

\author{Kohei Yamamoto,\authormark{1,2,3,*} and Ryan DeRosa\authormark{2}}

\address{
\authormark{1}Center for Space Sciences and Technology, University of Maryland, Baltimore County, 1000 Hilltop Circle, Baltimore, MD 21250, USA
\\
\authormark{2}NASA Goddard Space Flight Center, 8800 Greenbelt Road, Greenbelt, MD 20771, USA
\\
\authormark{3}Center for Research and Exploration in Space Science and Technology, NASA/GSFC, 8800 Greenbelt Road, Greenbelt, MD 20771, USA
}

\email{\authormark{*}y9m9k0h@gmail.com}

\begin{abstract*}
Simulating the propagation of a large beam with high spatial resolution is a common challenge in optics due to both computational time and memory demands.
While computational time sets a practical limit on what one can simulate, memory resources are an absolute hard limit.
In this paper, we report a novel method to overcome a gap in spatial scale between the beam and the relatively small optical disturbance that needs to be modeled, such as those due to particulate contamination.
The method splits the entire propagation chain into to two paths: a specular field propagation on coarse grids and scattered fields on fine grids.
Especially for the latter, we dynamically apply high-resolution local patches for individual particles.
Not limited to particulate contamination, this method is applicable to any tiny discontinuous structures on the beam propagation path without memory overflow triggered.
\end{abstract*}

\section{Introduction}
A general difficulty encountered in optical simulation is the range of spatial scales relevant to the optical system, including beams and defects, such as particulate contamination, that could adversely affect imaging or other measurements.
The large scale gap na\"{i}vely requires at the same time a relatively huge number of two-dimensional (2D) grid elements ($K$ and $L$) with a relatively tiny physical size for each element($dx$ and $dy$).
This significantly slows down numerical simulation via generic Fourier transform methods, or in an extreme environment, the massive required memory exceeds the available computational resources and triggers overflow.
Nevertheless, simulating the scattering of light into a large optical field by tiny defects will be important for future space missions that will be equipped with large mirrors and require extremely high contrast or signal-to-noise ratio, e.g., Habitable World Observatory (HWO)~\cite{HWO} or the Laser Interferometer Space Antenna (LISA)~\cite{LISA}.

Researchers have explored for decades many numerical techniques to overcome such computational difficulties that one encounters when simulating the real world.
Starting with a fast Fourier transform (FFT), and angular spectrum method (ASM) based on it, ~\cite{Cooley1965,goodman1968,Delen1998,Matsushima2009}, researchers have pursued other algorithms achieving Fourier transforms more efficiently for specific problems, such as the matrix triple product (MTP)~\cite{Soummer2007,Guizar-Sicairos2008} or the chirp z-transform (CZT)~\cite{Rabiner1969,Bluestein1970}.
The comparison of these methods for two-dimensional Fourier transforms in optical simulation is well summarized in~\cite{Jurling2018}.
At the same time, there is a large body of literature discussing the optical consequences of particulate contamination, including reference texts such as~\cite{BohrenHuffman1983} and~\cite{Fest2013}.
There is also a growing body of work investigating the consequences of micrometeoroid impacts~\cite{Gialluca2026}, which similarly may demand fine-resolution modeling of the scattering defects.

Despite all of the progress in efficiently simulating optical systems, there are some applications where the computational challenges provide an opportunity to develop novel methods.
While the extreme contrast goals of future coronagraphs have spurred an extended and intense modeling effort, including the effect of particles~\cite{Balasubramanian2009}, they are often concerned with the relatively larger bits of contamination, due to the angular distribution of scatter.
Conversely, inter-satellite laser communication links or distance measuring constellations like LISA must be concerned with the effects of coherent backscatter from the transmitter, which involves quantifying the effect of micron-level defects in significantly broader beams, stressing the computational problems inherent in these kinds of calculations.
In this paper, building on the existing algorithms, we propose a novel method to simulate an optical system having a great gap in spatial scale below.
The method enables us to compute, e.g., a meter-scale 2D optical surface contaminated by micro-scale particles without a memory overflow, which, if untreated, sets an absolute hard limit on what can be simulated.
After describing the method and discussing the computation cost, we show an example of the application to the propagation of a large Gaussian beam through a diffuser contaminated by numerous particles.
Although the precise modeling of scattering physics over the entire three-dimensional angle requires more complex mathematics such as Mie scattering, this work adopts a pure 2D phase mask to simplify the problems.

\section{Superposition}
In Fourier optics, based on the thin-element approximation, one can simulate the effect of optical surfaces on an incident beam using a phase mask, with phase $\phi$, written as 
\begin{align}
    E_\phi[k,l] &= e^{i\phi[k,l]}\cdot E_\mrm{in}[k,l],
    \label{eq:Eout1}
\end{align}
where $E_\mrm{in}$ and $E_\phi$ are complex optical fields right before and right after the optical surface.
Although we could use the non-unity amplitude for the mask to properly simulate effects such as absorption, we represent any effect of an optic and its imperfection by pure phase masks in this paper.
Throughout this paper, following the notation in~\cite{Jurling2018}, we denote coordinate grids using a capital letter ($X$) for the total number of elements in an axis and a lowercase ($x\in [0,X-1]$) for a specific element along that axis.

Importantly, \cref{eq:Eout1} can easily be decomposed into two parts: affected and unaffected fields:
\begin{align}
    E_\phi[k,l] &= E_\mrm{in}[k,l] + (e^{i\phi[k,l]}-1)\cdot E_\mrm{in}[k,l].
    \label{eq:Eout2}
\end{align}
The Lyot-style coronagraphs make use of this decomposition to apply different types of propagation methods to improve computational efficiency in both speed and memory consumption~\cite{Soummer2007}.

Our method could be viewed as a more flexible version of the one developed for the Lyot coronagraph.
Let us discuss, e.g., a transmissive surface contaminated by $P$ particles, each of which has the phase $\phi_p$ ($p\in [1,P]$).
In this case, the total phase $\phi$ in~\cref{eq:Eout2} over the $K\text{-}L$ plane can be written by
\begin{align}
    \phi[k,l] = \sum_{p\in P}\phi_p[k,l],
    \label{eq:phi}
\end{align}
where, for simplicity, we assumed no complexity caused by a rare case such as particle overlap~\footnote{Our method should be applicable to the case where some particles overlap by analyzing them together instead separately; however, throughout the rest of the paper, we ignore the overlap.}.
For later, we introduce the minimal information, which can be generated as a particle's metadata according to user-defined random distributions: ($x_{0,p}$, $y_{0,p}$) is the exact (i.e. not rounded to the spatial grid) global coordinate of the $p$-th particle center and $r_p$ is the radius of the $p$-th particle.
Let us also define the \emph{scattering phase mask} of individual particles as
\begin{align}
    \delta_p[k,l] = e^{i\phi_p[k,l]} - 1,
    \label{eq:delta_n}
\end{align}
which gives zero at a grid where $\phi_p[k,l]=0$, namely the region where no part of the $p$-th particle lies.

With these quantities, \cref{eq:Eout2} turns to the following:
\begin{align}
    E_\phi[k,l] &= E_\mrm{in}[k,l] + (e^{i\phi[k,l]}-1)\cdot E_\mrm{in}[k,l]
    \nonumber\\
    &= E_\mrm{in}[k,l] + (\prod_{p\in P}e^{i\phi_p[k,l]}-1)\cdot E_\mrm{in}[k,l]
    \nonumber\\
    &= E_\mrm{in}[k,l] + (\sum_{p\in P}\delta_p+\sum_{p<q}\delta_p\delta_Q+...+\prod_{p\in P}\delta_p)\cdot E_\mrm{in}[k,l].
    \label{eq:Eout_par}
\end{align}
As we assume that particles never overlap with each other (i.e., only one of $\delta_p$ becomes non-zero at a certain single point in space), most of the scattering terms vanish and the equation significantly reduces to
\begin{align}
    E_\phi[k,l] &= E_\mrm{in}[k,l] + \sum_{p\in P}\delta_p[k,l]\cdot E_\mrm{in}[k,l].
    \label{eq:Eout_par2}
\end{align}
We can also incorporate a background phase $\phi_b[k,l]$ representing an optic, such as a lens, to simulate a contaminated optical component:
\begin{align}
    E_\phi[k,l] &= e^{i\phi_b[k,l]}(E_\mrm{in}[k,l] + \sum_{p\in P}\delta_p[k,l]\cdot E_\mrm{in}[k,l])
    \nonumber\\
    &= E^b_\phi[k,l] + \sum_{p\in P}E^p_\phi[k,l],
    \label{eq:Eout_par2b}
\end{align}
where we define $E^b_\phi = e^{i\phi_b[k,l]}E_\mrm{in}$, and $E^p_\phi[k,l]\coloneqq \delta_p[k,l]\cdot E^b_\phi[k,l]$.

Because of this linearity, without the violation of energy conservation, we can separately simulate the interaction with particles, individually propagate fields by ASM, and superimpose all light fields, including the non-scattered specular field, on an output plane:
\begin{align}
    E_\mrm{out}[k',l'] &= \mrm{ASM}[E_\phi[k,l]]
    \nonumber\\
    &= \mrm{ASM}[E^b_\phi[k,l]] + \mrm{ASM}[\sum_{p\in P}E^p_\phi[k,l]]
    \nonumber\\
    &= \mrm{ASM}[E^b_\phi[k,l]] + \sum_{p\in P}\mrm{ASM}[E^p_\phi[k,l]]
    \nonumber\\
    &= E^0_\mrm{out}[k',l'] + \sum_{p\in P}E^p_\mrm{out}[k',l'].
    \label{eq:E_final_asm}
\end{align}
We define $E^0_\mrm{out}[k',l']\coloneqq\mrm{ASM}[E^b_\phi[k,l]]$ and $E^p_\mrm{out}[k',l']\coloneqq\mrm{ASM}[E^p_\phi[k,l]]$, where $\mrm{ASM}[x]$ represents an arbitrary ASM operation.
For generality, we assumed that the global $K\text{-}L$ input plane is transformed to the global $K'\text{-}L'$ output plane.

\section{Off-axis ASM}
We provide the formulation of the off-axis (or ``shifted") ASM~\cite{Matsushima2010}, which, with the superposition above, is the core for our method.
Instead of having the input and output planes share the same sample spacing and grid sizes, we assume that they can be independently chosen, as we will implement the off-axis ASM based on CZT.

\begin{figure}[h]
    \centering
    \includegraphics[width=\linewidth]{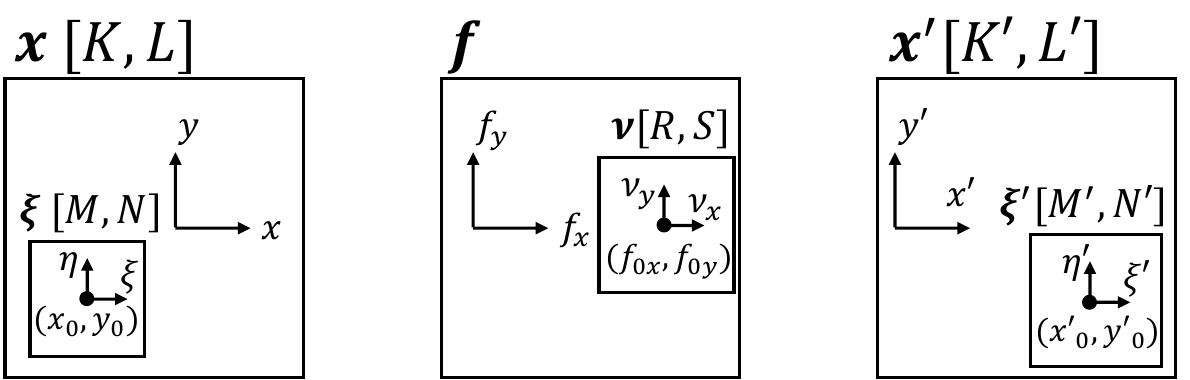}
    \caption{
    Coordinate system for the formulation of the off-axis ASM.
    $\bm{x}$, $\bm{f}$, and $\bm{x'}$ are the global input spatial, spatial frequency, and output spatial planes.
    $\bm{\xi}$, $\bm{\nu}$, and $\bm{\xi'}$ are the corresponding local patches.
    Each plane/patch ($\bm{x}$, $\bm{\xi}$, $\bm{\nu}$, $\bm{x'}$, and $\bm{\xi'}$) has the grid size shown in the brackets.
    The resolutions are represented by $d(coordinate)$ (e.g., ($d\xi$, $d\eta$)), and the physical total ranges are represented by $D(coordinate)$ (e.g., ($D\xi'$, $D\eta'$)).
    }
    \label{fig:tilted_asm_coord}
\end{figure}

As shown in \cref{fig:tilted_asm_coord}, we denote the global spatial coordinates of the input and output by $\bm{x}=(x, y)$ and $\bm{x'}=(x', y')$, and the local coordinate centered at $\bm{x_0}$ and $\bm{x_0'}$ by $\bm{\xi}=\bm{x}-\bm{x_0}=(\xi, \eta)$ and $\bm{\xi'}=\bm{x'}-\bm{x_0'}=(\xi', \eta')$.
The Fourier space is also denoted in a similar manner: the global frequency coordinate is $\bm{f}=(f_x, f_y)$, and the local one corresponding to the chosen carrier frequency $\bm{f_0}=(f_{0x}, f_{0y})$ is $\bm{\nu}=\bm{f}-\bm{f_0}=(\nu_x, \nu_y)$.
The resolutions are represented by $d(coordinate)$, e.g,. $d\xi$ and $d\eta$ for the local input patch.

Let us examine the propagation of $E(\bm{x})$ inside the local patch $\bm{\xi}$, i.e., $E(\bm{x_0+\xi})$.
If we choose the angular frequency $\bm{f_0}$ as the central frequency of interest (the ``carrier"), the carrier-relative field $U(\bm{\xi})$ is written by
\begin{align}
    U(\bm{\xi}) &= E(\bm{x_0+\xi})\cdot \mrm{e}^{-i2\pi\bm{f_0}(\bm{x_0+\xi})}.
    \label{eq:Uxi}
\end{align}
Hence, their Fourier transforms $\hat{E}(\bm{f_0+\nu})$ and $\hat{U}(\bm{\nu})$ are related by
\begin{align}
    \hat{E}(\bm{f_0+\nu}) &= \iint E(\bm{x_0+\xi})\cdot \mrm{e}^{-i2\pi(\bm{f_0} + \bm{\nu})(\bm{x_0+\xi}) } d\bm{\xi}
    \nonumber\\
    &= \mrm{e}^{-i2\pi\bm{\nu}\bm{x_0}} \cdot \iint U(\bm{\xi})\cdot \mrm{e}^{-i2\pi\bm{\nu}\bm{\xi}} d\bm{\xi}
    \nonumber\\
    &= \mrm{e}^{-i2\pi\bm{\nu}\bm{x_0}} \cdot \hat{U}(\bm{\nu}).
    \label{eq:Ehat_Uhat}
\end{align}

In general, the ASM propagates the field from the input plane to the output plane with the forward and inverse Fourier transforms as,
\begin{align}
    E'(\bm{x_0'+\xi'}) &= \mrm{ASM}[E(\bm{x_0+\xi})]
    \nonumber\\
    &= \iint H(z; k_z) \hat{E}(\bm{f_0+\nu}) \cdot \mrm{e}^{i2\pi(\bm{f_0+\nu})(\bm{x_0'+\xi'}) } d\bm{\nu},
    \label{eq:Eprime}
\end{align}
having a propagator $H(z; k_z)$ defined as
\begin{align}
    H(z; k_z) &= \mrm{e}^{i k_z z},
    \label{eq:Hz}\\
    k_z &= \sqrt{k^2 - k_x^2 - k_y^2},
    \label{eq:kz}
\end{align}
where $(k_x,k_y)=(2\pi (f_{0x}+\nu_x), 2\pi (f_{0y}+\nu_y))$.
Note that the wave vector $\bm{k}$ here is not in the local coordinate, but evaluated in the global coordinate.

Using \cref{eq:Uxi} for the output coordinate, the combination of \cref{eq:Ehat_Uhat,eq:Eprime} gives us
\begin{align}
    U'(\bm{\xi'}) &= E'(\bm{x_0'+\xi'})\cdot \mrm{e}^{-i2\pi\bm{f_0}(\bm{x_0'+\xi'})}
    \nonumber\\
    &= \mrm{e}^{-i2\pi\bm{f_0}(\bm{x_0'+\xi'})} \cdot \iint H(z; k_z) \hat{E}(\bm{f_0+\nu}) \cdot \mrm{e}^{i2\pi(\bm{f_0+\nu})(\bm{x_0'+\xi'}) } d\bm{\nu}
    \nonumber\\
    &= \iint H(z; k_z) \mrm{e}^{i2\pi\bm{\nu}(\bm{x_0'} - \bm{x_0}) } \cdot\left(\iint U(\bm{\xi})\cdot \mrm{e}^{-i2\pi\bm{\nu}\bm{\xi}} d\bm{\xi}\right) \cdot \mrm{e}^{i2\pi\bm{\nu}\bm{\xi'} } d\bm{\nu}.
    \label{eq:Uprime}
\end{align}
This tells us how to propagate the carrier-relative field within the input small patch $U(\bm{\xi})$ to the carrier-relative field within the output small patch $U'(\bm{\xi'})$.

Lastly, as obvious from the first line of \cref{eq:Uprime}, we can easily recover the absolute field in the global coordinate of the output plane $\bm{x'}$:
\begin{align}
    E'(\bm{x_0'+\xi'}) &= U'(\bm{\xi'}) \cdot \mrm{e}^{i2\pi\bm{f_0}(\bm{x_0'+\xi'})},
    \label{eq:E'x'}
\end{align}
which is an essential step when adding the contribution of individual particles to the background field, as shown in \cref{eq:E_final_asm}.

The protocol in this section could be summarized by the following three steps:
\begin{enumerate}
    \item Flattening in \cref{eq:Uxi}: We isolate the input carrier-relative field $U(\bm{\xi})$ to the chosen carrier frequency $\bm{f_0}$.
    \item Propagation in \cref{eq:Uprime}: We propagate $U(\bm{\xi})$ to the output small patch $\bm{\xi'}$, considering the difference in the global centers $\bm{x_0'} - \bm{x_0}$.
    \item Unflattening in \cref{eq:E'x'}: We recover the absolute field in the global coordinate of the output plane $E'(\bm{x_0'+\xi'})$.
\end{enumerate}

\section{Implementation}\label{sec:implementaion}
Our method separately propagates all terms in \cref{eq:Eout_par2b}, as shown in \cref{eq:E_final_asm}.
We can use well-established techniques of Fourier transforms for the specular field on the coarse grids such as DFT or CZT depending on the global input plane $\bm{x}$ and the global output plane $\bm{x'}$.
Meanwhile, we apply the small patches $\bm{\xi}$ to individual particles, propagate the scattered lights using the CZT-based off-axis ASM to the local output patch $\bm{\xi'}$, and rasterize them to the global output plane $\bm{x'}$.
Arbitrarily sampled Fourier methods such as the CZT allow for an analysis which utilizing appropriate resolutions for the particles in the input plane while maintaining common sampling in Fourier space for the section of field angles which are of interest.
This processing can be fully parallelized between particles.
In the following, we discuss some crucial steps specific for the implementation of our method.
Although the notation of the coordinates basically follows \cref{fig:tilted_asm_coord}, the patches gain subscript $p$, as they depend on the global center of the particles $(x_{0,p},y_{0,p})$ and the radius $r_p$.
As a general remark, all coordinates must satisfy Nyquist sampling constraints.

First, the local input patches can be centered at the exact center of individual particles: $\bm{x_0}=\bm{x_{0,p}}=(x_{0,p},y_{0,p})$.
As the local-patch resolutions $(d\xi,d\eta)$ are normally not equal to, but smaller (i.e. finer), than the global resolutions $(dx,dy)$ for problems that require our method, there is no grid to which we need to rasterize the particle's center $(x_{0,p},y_{0,p})$.
The grid sizes $(M_p,N_p)$ would not need extra zero padding around a particle with $r_p$, as the interacted field $E^p_\phi[m_p,n_p]$ is basically guaranteed to be zero outside the particle and to be unaffected by artificial Gibbs ringing by the definition of $\delta_p$.

Second, we need to define the spatial frequencies of the carrier $\bm{f_{0,p}}=(f_{0x,p},f_{0y,p})$, and also the (possibly common) frequency range $(D \nu_x, D \nu_y)$.
The quantities depend on problems to tackle; E.g., if one wants to track the scattered light around the specular field bent by a lens, we would need to calculate the bending angle of the lens in $\bm{\xi_p}$, e.g., $(\theta_{x,p},\theta_{y,p})$, and convert it to the frequency via $f=\sin(\theta)/\lambda$.

Third, with respect to the local output patch $\bm{\xi'_p}$, this needs to be conveniently defined to superimpose all lights on the global output plane $\bm{x'}$.
To begin with, the center of the local output patch $\bm{x_{0,p}'}=(x_{0,p}',y_{0,p}')$ can be determined by $\bm{x_0'}$, $\bm{f_{0,p}}$, and the propagation distance from the scatterer $Dz$ according to ray representation.
In addition, we derive the sizes of the output grid $(M',N')$ from $(D \nu_x,D \nu_y)$ and $Dz$ in such a way that all frequency components fall onto the output patch.
Note that we use the common size $(M',N')$, assuming that we apply the same frequency range of interest to particles and the spread of light dominates the effect of the input patch size $(D \xi_p, D \eta_p)$ on the required output patch size; hence, $\bm{\xi_p'}\rightarrow\bm{\xi'}$.
Importantly, unlike the center of the local input patch $\bm{x_{0,p}}$ discussed above, $\bm{x_{0,p}'}$ cannot be the exact values.
In order to superimpose all lights, as shown in \cref{eq:E_final_asm}, we need to add the local output patch to the global output plane;  $E'(\bm{x_0'+\xi'})\rightarrow E'(\bm{x'})$.
Hence, we should snap the output center $\bm{x_{0,p}'}$ to the global lattice, namely $[\bm{x_{0,p}'}/d\bm{x'}]\cdot d\bm{x'}$, and then define the patch around it.
Note that $d\bm{\xi'}$ should always be equal to $d\bm{x'}$.

Fourth, regarding the Fourier space $\bm{\nu_p}$, one determines the resolution $(d\nu_x, d\nu_y)$ to meet the Nyquist sampling: $(d\nu_x, d\nu_y)<(1/(M'dx'), 1/(N'dy'))$.
After that, the grid size $(R, S)$ is naturally determined by the resolutions and the physical range $(D \nu_x, D \nu_y)$.
We also assumed that the relative Fourier space is common for the particles: $\bm{\nu_p'}\rightarrow\bm{\nu'}$.
To avoid artificial Gibbs ringing, one may apply a anti-aliasing filter in the local Fourier space.

Fifth, concerning the CZT-based propagation core in \cref{eq:Uprime}, it is expressed based on the carrier-relative fields ($U_p(\bm{\xi_p})$ and $U_p'(\bm{\xi'})$).
This means that any coordinates required for the propagation ($\bm{\xi_p}$, $\bm{\nu}$, and $\bm{\xi'}$) are zero-centered.
This in turn provides the capability to precompute the CZT's ingredients ($a$, $b$, and $\hat{H}$ in~\cite{Jurling2018}), determined only by the grid sizes, as common values for a group of particles.
As $(R,S)$ and $(M',N')$ are assumed to be common for all particles, a rational way of classifying particles would be to compute percentiles of their radius.
For particles in the same percentile (e.g., $q$-th percentile), we could apply the same grid size of the local input patch $(M_q,N_q)$, and then compute the common CZT's ingredients.
This prevents us from computing the CZT's ingredients (especially $\hat{H}$ in~\cite{Jurling2018}, which includes 2D Fourier transform) for every particle.

Lastly, to wrap up this section, we note the computational cost of our method.
The asymptotic complexity of ASM based on the standard DFT $t_\mrm{DFT}$ is given by
\begin{align}
    t_\mrm{DFT} \propto 2 KL \log_2(KL),
    \label{eq:tDFT}
\end{align}
where \num{2} represents a pair of forward and inverse Fourier transforms as required by ASM, and we assume $(K,L)=(K',L')$ as required by DFT.
On the other hand, to represent the asymptotic complexity of ASM based on our method $t_\mrm{patch}$, we introduce four integers $\hat{K}_q$, $\hat{L}_q$, $\hat{K}'$, and $\hat{L}'$, as required by CZT:
\begin{align}
    \hat{K}_q &\ge M_q + R -1,
    \label{eq:Khat}\\
    \hat{L}_q &\ge N_q + S -1,
    \label{eq:Lhat}\\
    \hat{K}' &\ge R + M' -1,
    \label{eq:Kprimehat}\\
    \hat{L}' &\ge S + N' -1.
    \label{eq:Lprimehat}
\end{align}
In practice, one would choose the minimum fast lengths for the internal DFT of CZT that meet these inequalities.
If we assume using standard DFT for the specular-light propagation, the asymptotic complexity of our method would be given by
\begin{align}
    t_\mrm{patch} &\propto 2 KL \log_2(KL) + \sum_q P_q\cdot(2\hat{K}_q\hat{L}_q\log_2(\hat{K}_q\hat{L}_q) + 2\hat{K}'\hat{L}'\log_2(\hat{K}'\hat{L}'))
    \nonumber\\
    &\approx 2 \sum_q P_q\cdot(\hat{K}_q\hat{L}_q\log_2(\hat{K}_q\hat{L}_q) + \hat{K}'\hat{L}'\log_2(\hat{K}'\hat{L}')),
    \label{eq:tpatch}
\end{align}
where $P_q$ represents the number of particles in the $q$-th percentile, and $\sum_q$ is the summation of all percentiles.
As being obvious from the comparison between \cref{eq:tDFT} and \cref{eq:tpatch}, for any computational setup DFT is capable of and where we use the common $(K,L)$ for both methods, our method is painfully slower than conventional DFT by simply adding the excess computational cost of the scattering components.
However, even if $(K,L)$ becomes much larger than available computational resources and DFT triggers memory overflow, our method provides an opportunity to significantly reduce $(K,L)$ by using the coarse grids for the specular component.
In that case where our method is useful, the scattered components dominate the computational cost, as shown in the second line of \cref{eq:tpatch}.
Likely, the inverse Fourier transforms scaled by $\hat{K}'\hat{L}'$ are dominant, as $(M',N')$ would need to be larger than $(M_q, N_q)$ to properly capture the diffracted light from the local input patch.

\section{Demonstration}
In this section, we conduct two types of demonstration: a verification cross-check of our method and the simulation of a massive beam in comparison to the applied particulate dimension.
For both demonstrations, we commonly used the Tukey window with $\alpha=2$ for the Fourier space.
Furthermore, the particle distribution follows CL500 as defined in IEST-STD-CC1246E~\cite{cc_spec}.
Based on the distribution and the area of the scatterer, we generate a list of particle metadata (the radius $r_p$ and the location ($x_{0,p}$, $y_{0,p}$)) with a uniform distribution for the location.
The number of generated particles can be obtained from the size of the list.

\begin{table}[htbp]
    \centering
    \caption{\bf
    Parameters for the verification and large beam demonstration.
    Some common parameters:
    The refractive index of the particles: $n=1.5$;
    Object position: $z=\SI{1}{\centi\meter}$.
    }
    \begin{tabular}{ccc}
    \hline
    Parameters & Cross-check & Large Beam \\
    \hline
    Geometry & Lens & Free-space  \\
    Detector position & $z=\SI{5}{\centi\meter}$ & $z=\SI{5}{\meter}$ \\
    Beam radius & \SI{0.5}{\milli\meter} & \SI{15}{\centi\meter}  \\
    Lens focal length & \SI{100}{\milli\meter} & N/A  \\
    Radius of an scatterer & \SI{4}{\milli\meter} & \SI{20}{\centi\meter}\\ 
    \# of particles & 185 & \num{5761774} \\
    ($Dx$, $Dy$) & (\SI{2}{mm}, \SI{2}{mm}) & (\SI{40}{cm}, \SI{40}{cm}) \\
    ($dx$, $dy$) & (\SI{1}{\micro\meter}, \SI{1}{\micro\meter}) & (\SI{1}{\milli\meter}, \SI{1}{\milli\meter}) \\
    ($d\xi$, $d\eta$) & ($dx$, $dy$) & (\SI{1}{\micro\meter}, \SI{1}{\micro\meter}) \\
    Angle range & $\pm$ \SI{50}{\milli\radian} & $\pm$ \SI{50}{\micro\radian}, $\pm$ \SI{150}{\micro\radian}, and $\pm$ \SI{250}{\micro\radian} \\
    ($df_x$, $df_y$) & ($1/Dx$, $1/Dy$) & ($1/(M'dx')$, $1/(N'dy')$) \\
    ($Dx'$, $Dy'$) & ($Dx$, $Dy$) & ($Dx$, $Dy$) \\
    ($dx'$, $dy'$) & ($dx$, $dy$) & ($dx$, $dy$) \\
    \hline
    \end{tabular}
    \label{tab:params}
\end{table}

First, we show that our method produces results consistent with the conventional DFT-based propagation.
As it is computationally not possible for conventional DFT to work on a massive beam with a super-fine resolution, the experimental setup must be of a comfortably small size, whose parameters are listed in \cref{tab:params}.
For comparison, our method also adopts almost the same set of parameters as DFT, except for the small input patches for individual particles.
Although this means that the output patch is not small but a global output plane itself (hence, $\bm{\xi'}=\bm{x'}$  and $\bm{x_{0,p}'}=(0,0)$), a separate investigation confirmed that using small output patches did not introduce more noise than numerical errors.

\begin{figure}[h]
    \centering
    \includegraphics[width=\linewidth]{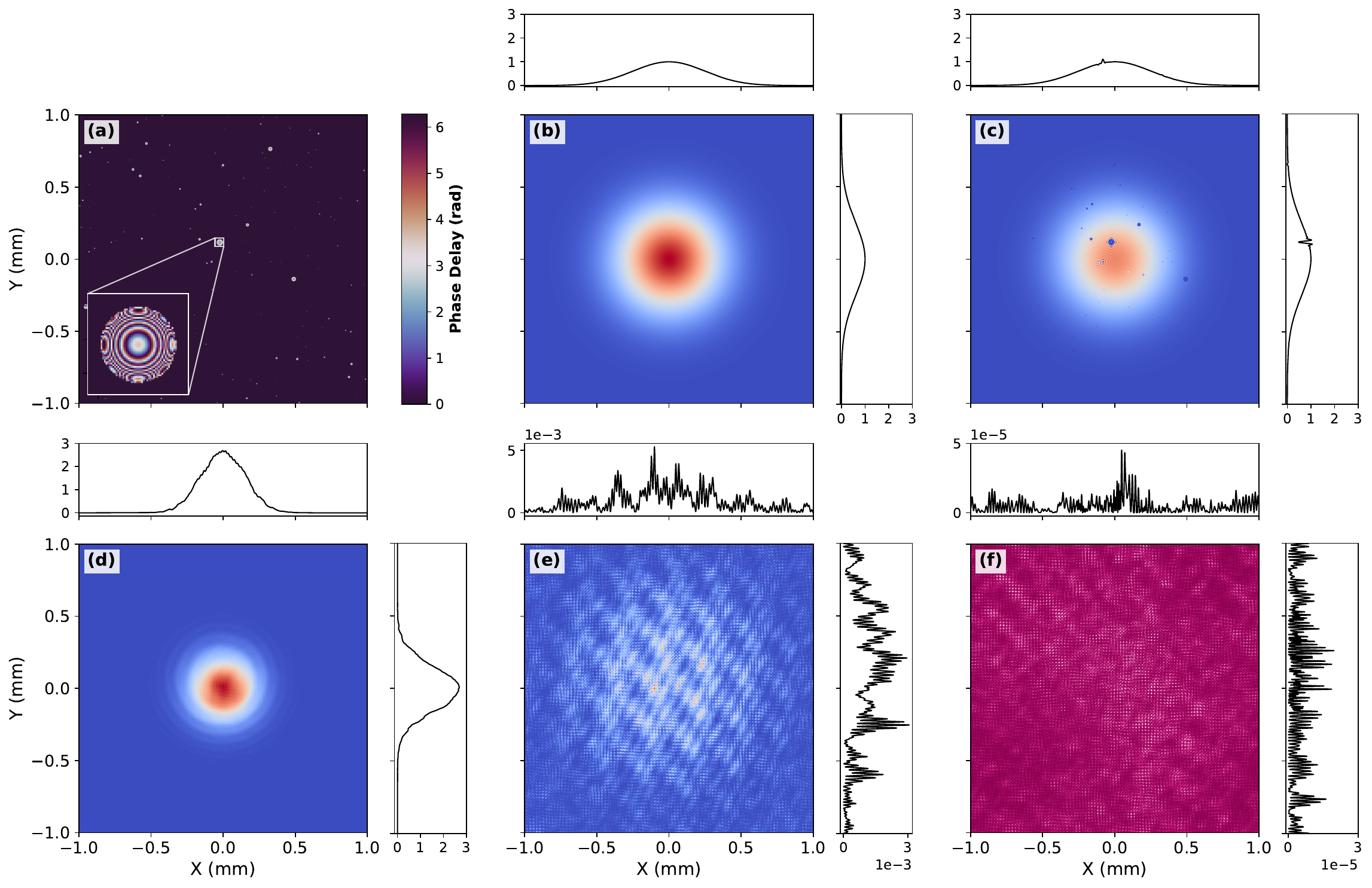}
    \caption{
    Results of the cross-check test.
    (a) is the phase map, wrapped by $[0,2\pi]$, showing the particle distribution over the $x$-$y$ plane.
    The inset graph zooms into the largest particle.
    (b)-(f) show the intensity at different longitudinal positions.
    (b) is the input beam at $z=\SI{0}{\meter}$.
    (c) is the beam right behind the contaminated lens at $z=\SI{1}{\centi\meter}$.
    (d) and (e) are the detected beam (the total and the scattered, respectively) computed by our local-patch method at $z=\SI{5}{\centi\meter}$.
    (f) is the intensity of its difference from the resulting field computed by conventional DFT.
    }
    \label{fig:cross_check}
\end{figure}

The results are summarized in \cref{fig:cross_check}.
The input beam at $z=\SI{0}{\meter}$ in \cref{fig:cross_check}b propagates with conventional DFT, and reaches a lens contaminated by particles in \cref{fig:cross_check}a.
Once the beam interacts with the contaminated lens, it is affected as shown in \cref{fig:cross_check}c.
In our local-patch method, we do not generate the interacted beam in the $K\text{-}L$ global coordinate at all; instead, the specular component is propagated to the detector plane with conventional DFT, while the small patches are assigned for the individual particles, and their propagation is simulated individually.
Finally, all components (i.e., the specular field and 185 scattered lights) are all superimposed on the detector plane at $z=\SI{5}{\centi\meter}$, forming the beam shown in \cref{fig:cross_check}d.
\cref{fig:cross_check}e shows only the scattered light computed in our method, which shows a peak height of around $5.0\cdot 10^{-3}$ for the cross section of $y'=0$.
Finally, the complex field on the detector plane in our method is subtracted from the one produced by the conventional DFT, whose intensity is shown in \cref{fig:cross_check}f.
This shows a peak height of around $4.5\cdot 10^{-5}$ for $y'=0$, and this is less than \SI{1}{\percent} of the aforementioned peak height of absolute speckles, so the agreement between methods is reasonably tight.

The dominant factor of the residual difference was identified as the rasterization of the particles in the DFT: a conventional DFT is forced to evaluate the particles in predefined fixed global grids, shown in \cref{fig:cross_check}a.
However, as discussed in the previous section, our method uses the exact particle locations and defines the individual local patches centered on them.
This mismatch in the input sampling lattices causes the phase differences through $k_x\delta x$ where $\delta x$ is up to the half of the pixel size $dx$, namely \SI{0.5}{\micro\meter}, affecting higher frequencies more.
Nevertheless, this is just the difference of the evaluated lattices and does not mean our method violates physics at this level.

Now that we have a satisfactory cross-check against conventional methods, let us move on to the demonstration of a massive beam contaminated by numerous small particles, which might be expected to be computationally prohibitive with a typical DFT approach.
The parameter set is also shown in \cref{tab:params}.
For example, we computed different physical angle ranges: $\pm\SI{50}{\micro\radian}$, $\pm\SI{150}{\micro\radian}$, and $\pm\SI{250}{\micro\radian}$, all much larger than the asymptotic angle of the beam, which is around \SI{2.3}{\micro\radian}.
Importantly, for individual ranges, we computed optimal parameters for the Fourier space $\bm{\nu}$ and the local output patch $\bm{\xi'}$:
As the inverse of the transition bandwidth of the Tukey window filtering the Fourier space, we derived the physical size of the spatial ringing, and added margin to the local output patch, due to the finite bandwidth $(\Delta\xi',\Delta\eta')$.
This gives $(M',N')$.
Based on $(\Delta\xi',\Delta\eta')$, we can derive $(df_x, df_y)$ via the Nyquist sampling, which in turn determines $(R,S)$.
This way, with the common $\alpha$ for the Tukey window, the broader angle range leads to the smaller $(M',N')$ (assuming the ringing dominates $\bm{\xi'}$) but to the larger $(R,S)$.
Hence, the computational cost is not monotonic against the angle range.

\begin{figure}[h]
    \centering
    \includegraphics[width=\linewidth]{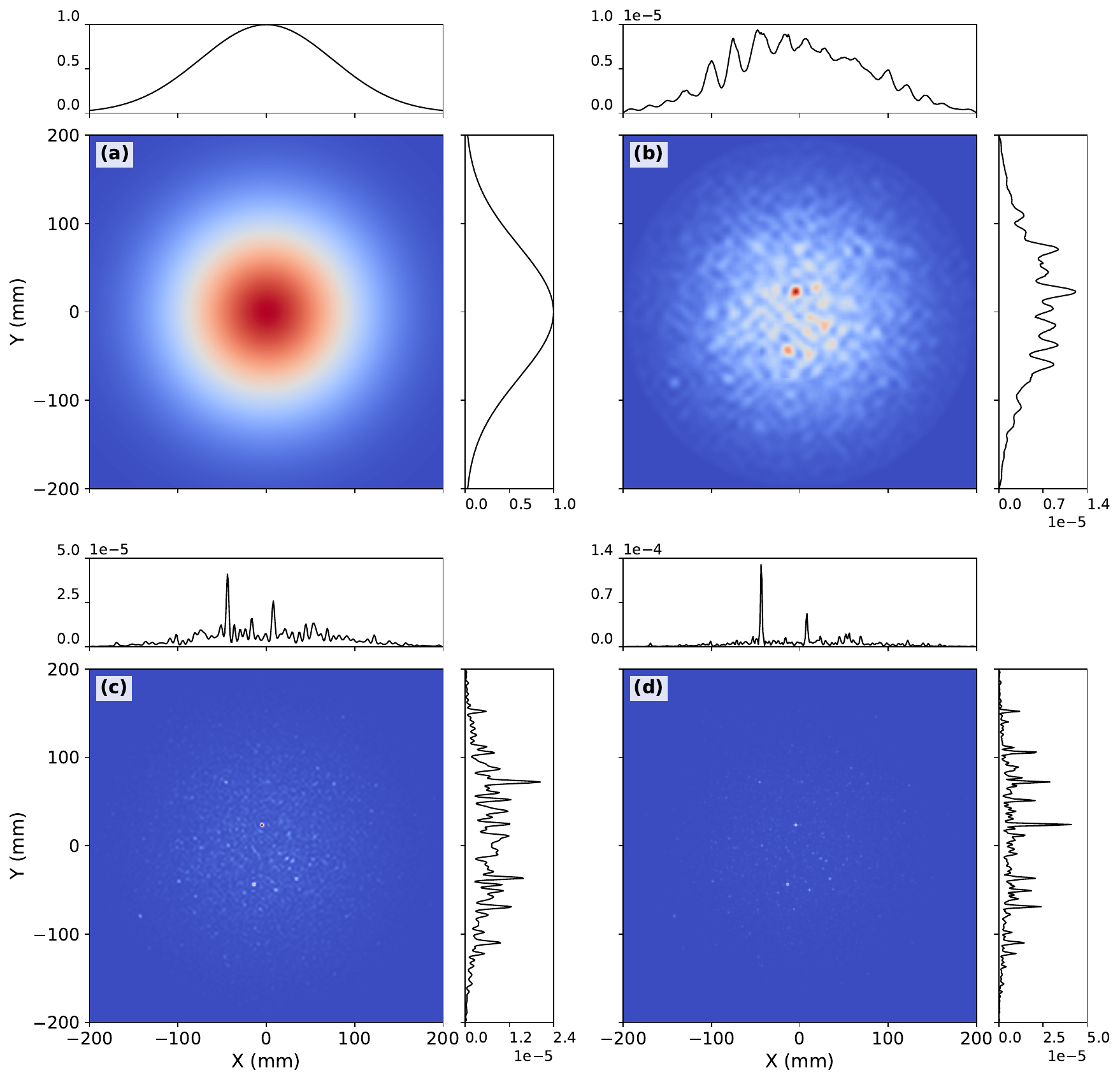}
    \caption{
    Results of the scattering analysis with a massive beam with the $1/\mrm{e}^2$ radius of \SI{15}{\centi\meter}.
    All plots are shown in intensity.
    (a) is the input beam at $z=\SI{0}{\meter}$.
    (b) - (d) show the scattered lights on the detector at $z=\SI{5}{\meter}$ with the different angular ranges: $\pm\SI{50}{\micro\radian}$ for (b), $\pm\SI{150}{\micro\radian}$ for (c), and $\pm\SI{250}{\micro\radian}$ for (d).
    }
    \label{fig:massive_run}
\end{figure}

The results are shown in \cref{fig:massive_run}.
Our method successfully completed the computation of scattered lights of the massive \SI{30}{cm}-diameter Gaussian beam by \num{5761774} particles, whose radius is down to \SI{1}{\micro\meter}, without memory overflow.
For the three angle ranges $\pm\SI{50}{\micro\radian}$ ($R=S=8$; $M'=N'=61$), $\pm\SI{150}{\micro\radian}$ ($R=S=22$; $M'=N'=60$) and $\pm\SI{250}{\micro\radian}$ ($R=S=24$; $M'=N'=39$), the computation times were \SI{202}{\second}, \SI{244}{\second}, and \SI{209}{\second}, respectively, with the 2023 MacBookPro with the Apple M2 Pro chip, \SI{16}{GB} memory, and 8 performance cores used for parallel computation over particles.
Relative computational times basically follow the order of asymptotic complexities when we substitute each pair of $R$, $S$, $M'$, and $N'$ for \cref{eq:tpatch}.

\section{Conclusion}
Numerical simulations in wave optics inherently face a trade-off between spatial accuracy and computational resources.
When determining the spatial resolution and domain size, computational time imposes a practical limit on what can be simulated.
In the meantime, memory capacity establishes an absolute hard limit on the scale of data that can be analyzed.
In conventional DFT, attempting to adequately sample a macroscopic spatial plane with microscopic resolution inevitably triggers memory overflow, rendering the computation impossible.

This paper presents a novel computational method designed to bridge the extreme scale gap between macroscopic optical beams (e.g., meter-scale) and microscopic discontinuous structures such as particulate contamination (e.g., micro-scale) without triggering memory overflow.
Rather than attempting to simulate a global field at an unfeasibly high resolution, our approach decomposes the beam propagation into its specular and scattered components, akin to techniques developed for Lyot-style coronagraphy.
The core novelty of this work lies in assigning dynamically scaled local spatial patches to individual particles.
These independent patches are propagated to local output patches using the CZT-based off-axis ASM, after which the resulting scattered fields are coherently superimposed onto the global output plane together with the specular component propagated in a conventional manner.
Provided that either of the individual global planes or local patches does not exceed hardware limits, this framework entirely circumvents the memory overflow bottleneck traditionally associated with high-ratio multiscale optical simulations.

As shown in \cref{fig:cross_check}, our method demonstrates excellent agreement with the conventional DFT approach, yielding a fractional difference in intensity of less than \SI{1}{\percent}.
This minor residual error can be directly attributed to input grid rasterization effects rather than any underlying physical inaccuracies in the propagation physics.
Furthermore, we successfully demonstrated the scalability of this technique by simulating the scattering effects of \num{5761774} particles—with radii as small as \SI{1}{\micro\meter}—within a \SI{30}{\centi\meter}-diameter beam.
This massive multiscale computation was executed without memory overflow and completed in a reasonable time frame of \SI{200}{\second} to \SI{250}{\second}.

Although we represent the scattering effects of particles by a pure phase mask in this study, future efforts should be invested to incorporate more accurate descriptions of scattering physics, e.g., Mie scattering, into the algorithm in a computationally efficient manner.

\smallskip

\begin{backmatter}
\bmsection{Funding}
The authors acknowledge the support from the NASA Physics of the Cosmos (PhysCOS) program.
K.Y.’s work is supported by NASA under Award No. 80GSFC24M0006.

\bmsection{Acknowledgment}
The authors thank Alden S. Jurling for useful discussions.

\bmsection{Disclosures} The authors declare no conflicts of interest.

\bmsection{Data Availability Statement}
Data underlying the results presented in this paper are not publicly available at this time but may be obtained from the authors upon reasonable request.

\end{backmatter}

\bibliography{references}

\end{document}